\documentstyle[aps,epsfig,pra]{revtex}  
  
\newcommand{\be}{\begin{eqnarray}}  
\newcommand{\ee}{\end{eqnarray}}

\def\ra{\rangle}

\begin{document}  
  
\title{Exact gate-sequences for universal quantum computation using the 
XY-interaction alone}  
  
\author{J. Kempe$^{1,2}$ and K.B. Whaley$^2$}  
\address {$^1$ CNRS-LRI, UMR 8623, Universit\'e de Paris-Sud, 91405 Orsay,  
France \\$^2$ Department of Chemistry, University of California, Berkeley }  
  
\date{\today}  
\maketitle  
  
\begin{abstract}  
In a previous publication \cite{Kempe:01a} we showed that it is possible to  
implement universal quantum computation with the anisotropic XY-Heisenberg  
exchange acting as a single interaction. To achieve this we used encodings of  
the states of the computation into a larger Hilbert space. This proof is non- 
constructive, however, and did not explicitly give the trade-offs in time that  
are required to implement encoded single qubit operations and encoded 
two-qubit gates. Here we explicitly give the gate-sequences needed to simulate 
these operations on encoded qubits and qutrits (three-level systems) 
and analyze the trade-offs involved. We also propose a possible layout for the qubits in a triangular arrangement. 
  
\end{abstract}  
  
\section{Introduction}  
  
Any quantum computation can be build out of simple operations involving only one  
or two qubits. These elementary quantum gates can generate any unitary operation  
over the qubits. One particularly simple set of universal gates consists of  
single qubit unitaries ($SU(2)$) together with an entangling two-qubit  
operation\cite{Nielsen:praise}.  
  
The physical implementation of these sets of gates in various proposed physical  
systems is often daunting, involving precise manipulations well beyond the  
current state of technology. Most physical systems, however, possess some  
intrinsic interactions that are easy to tune and to control. These interactions  
{\em per se} usually do not constitute a universal set of gates, in the sense  
that they cannot generate any arbitrary unitary transformation on the set of  
qubits (or qu{\em dits} - $d$-level systems). Thus, they generally have to be supplemented with  
additional means of applying the other interactions that are required in order  
to complete a universal gate set. It is the need to add this capacity that  
dramatically increases the device complexity and that may significantly diminish  
the decoherence times of the resulting quantum device, posing the largest  
challenge on the route to scalable universal quantum computation.   
  
It seems to be nearly a rule of thumb that one of these two kinds of  
interactions (single qubit/ two-qubit) is generally easy to achieve, whereas the  
other one is extremely hard to implement. An example of this easy/hard duality  
are the proposals based on solid state physics using quantum dots  
\cite{LossDiVincenzo:98,LossBurkard:99}, donor-atom nuclear spins \cite{Kane:98}  
or electron spins \cite{Vrijen}. In these approaches, the basic two-qubit  
quantum gate is generated by a tunable Heisenberg interaction (the Hamiltonian  
is $H_{ij}=J(t){\vec S}_i\cdot{\vec S}_j$ between spins $i$ and $j$), while the  
one-qubit gates require the  
control of a local Zeeman field.  Compared to the Heisenberg  
operation, the one-qubit operations are significantly slower and  
require substantially greater materials and device complexity, which  
may then also contribute to increasing the decoherence rate. Examples where  
single qubit gates are relatively easy to achieve whereas two-qubit gates are  
now hard, are the proposed quantum-optics implementations of quantum computers.  
Here quantum bits are implemented with two optical modes containing one photon  
\cite{Milburn:88}. In this setting, single qubit operations can be achieved with  
linear optics: beam-splitters and phase-shifters are sufficient to generate  
every rotation in $SU(2)$. Proposals for two-qubit gates, however, either suffer  
from the requirement for non-linear couplings between the optical  
modes\cite{Milburn:88}, or have to exploit feedback from photo-detectors. Such  
feedback results either in making the scheme non-deterministic (the coupling  
gate is merely probabilistic) or in requiring non-deterministic generation of  
entangled photon states\cite{Knill:01a}. Another recent proposal  
\cite{Ralph:01b} uses coherent states to encode an (approximate) qubit, which  
allows for implementation of an (approximate) controlled phase with just one  
beam-splitter but now relegates all the difficulty back again into the  
implementation of single-qubit interactions.  
  
Recently we have shown  
\cite{Kempe:01a,DiVincenzo:00a,Bacon:01,Kempe:ph,Bacon:PHD}  how to overcome  
this problem in several cases of interest. By suitably encoding the quantum  
states into a higher dimensional quantum system various interactions can be made  
universal. We have termed this method {\em encoded universality}  
\cite{Kempe:praise}. The Heisenberg interaction, for example, allows for encoded  
universality \cite{Kempe:00a} and we have identified the possible encodings to  
realize this. Our proof relies on Lie-algebraic methods and does not answer the  
very practical question: What price must be paid in additional gates to {\em  
implement} encoded single qubit and two-qubit operations? In  
\cite{DiVincenzo:00a} we have assessed the trade-off for the Heisenberg  
interaction. For the encoding of ``logical'' qubits into blocks of three  
``physical'' qubits each, we found that $4$ exchange gates in sequence allow to  
simulate every single-qubit operation on the encoded qubit, whereas a sequence  
of $19$ exchange gates gives the encoded controlled-NOT (up to single qubit  
operations). This was in serial mode: a parallel mode solution was also found  
that required only $3$ exchange gates for single-qubit operations and $7$  
exchange gates for the encoded controlled-NOT.  These sequences were obtained  
via numerical minimization of functions based on local invariants due to Makhlin  \cite{Makhlin:00}.  
  
Since then new results on encoded universality have been derived. In  
\cite{Bacon:01,Kempe:01a} we showed that the anisotropic exchange interaction  
lends itself to encoded universality and identified the required encoding to  
implement this. Results were obtained for general numbers of qubits $n$, with  
the smallest encoding required to achieve universality without additional  
interactions being an encoded qutrit (three-level system) constructed from $n=3$ physical qubits.   
In \cite{Wu:01a,Lidar:01a}, Lidar and Wu have analysed the universality  
properties of exchange-like interactions in the presence of additional single  
qubit $\sigma_z$ operations or static  $\sigma_z$ terms in the Hamiltonian. They  found that with the help of these additional $\sigma_z$ interactions it was  
possible to encode into pairs of qubits.  
  
Anisotropic Heisenberg spin couplings arise whenever there is some   
preferred direction in space along which the coupling is stronger or   
weaker. This could be due to, e.g., asymmetries induced by donor atoms   
in solid-state arrays of atoms coupled via their nuclear spins   
\cite{Kane:98,Kane:00}.  The XY-interaction arises when there is no   
coupling in the $z$-direction of the spins, while the coupling in $x$-   
and $y$-direction is equally strong:   
\begin{eqnarray}   
(H_{XY})_{ij}  = \frac{J_{ij}}{2} ({  \sigma}_x^i {  \sigma}_x^j+{   
\sigma}_y^i {  \sigma}_y^j) \equiv J_{ij} A_{ij}. \label{eq:H_xy}   
\end{eqnarray}   
This situation is relevant to several proposals for
solid-state quantum computation, {\em e.g.},    
using quantum dot spins and cavity QED,\cite{Imamoglu:99} and using nuclear
spins coupled by a two-dimensional electron gas \cite{Mozyrsky:01}.   
  
In the following we give an explicit assessment of the trade-offs involved in  
implementing universal computation with the XY-interaction alone. We give an explicit application to the $n=3$ qutrit encoding established by Lie group methods in  
Ref.~\cite{Kempe:01a} that possesses the smallest spatial overhead for logical 
encoding.  Consideration of possible gate sequences for this  
encoding shows that we may achieve an additional economy by using only two of  
the qutrit states to define a truncated qubit, {\em i.e.}, two encoded 
logical states from three physical qubits. In this case the {\em effective} encoding of the truncated qubit requires only two physical qubits and single qubit operations will require an additional ancillary qubit, which can be reused after the gate-application for subsequent gates.
We find that universal computation is possible on this truncated qubit using  
at most 7 XY exchange gates for encoded single qubit operations, and at most  
5 XY exchange gates for an encoded two-qubit gate  
(the controlled phase-flip ${C_z}$).  For the full qutrit, {\em i.e.}, 
using all three encoded states derived from the $n=3$ physical qubits, 
we find that 
single qubit operations may be constructed by slightly modifying the encoded single qubit operations used for the truncated qubit, which results in $12$-operations as a basis for single-qutrit operation while encoded two-qubit gates now require $8$ XY exchange gates.   

Thus the XY-anisotropic exchange interaction can be used to implement universal  quantum computation in an economical fashion without requiring supplemental  
interactions, whether static or dynamic.  This result is extremely attractive  
for experimental investigations, allowing application to encoding over arrays of degenerate quantum bits such as would arise from nanoscale fabrication methods,  {\i.e.}, with no requirement of a distinct energy spectrum.  
  
\section{The encodings}  
  
We will encode a qubit or qudit into parts of the Hilbert space of a system of  
$n$ qubits. As shown in \cite{Kempe:01a} the XY-interaction preserves the number  
of zeros and ones of a computational basis state. Within a space spanned by  
states with a fixed number of zeros and ones the XY-interaction is universal. We  
can chose a subspace spanned by states with $i$ ones and $n-i$ zeros (such that  
$i \leq (n-i)$) to encode ${{n}\choose{i}} =d$ basis states, yielding an encoded  
qu{\em dit}. The smallest encoded qudit is a qutrit encoded into $n=3$ qubits,  
according to  
\be  
\label{eq:qutrit}  
|0_L\ra=|100\ra \quad |1_L\ra=|010\ra \quad |2_L\ra=|001\ra.  
\ee  
For large $n$, the encoding efficiency (number of encoded qubits over number of  
physical qubits $=\log_2 d/n$) approaches unity.   
  
Quantum circuits built from encoded states have to respect the tensor product  
structure of the quantum circuit model.  
To map the encoded Hilbert space to the quantum circuit model, a cut-off has to  
be chosen ($n$ physical qubits encoding $d$ states).  
Blocks of $n$ qubits will represent the encoded qu{\em dits}. Since the tensor  
product of two of these blocks  is still spanned by basis-states with a constant  
ratio of zeros and ones (in a space of 2n qubits now), the product state is  
immersed in a subspace over which the XY-interaction is universal. In particular  
any two qudit gate can be implemented with XY-interactions. If $d$ is not a  
power of $2$ for a given encoding of $d$ states into $n$ qubits, we can always  
chose to not use some of the $d$ states, so that the remaining $2^k$ states form  
the $k$ logical qubits.   
  
The basic units of quantum circuits in the implementation of most algorithms are  
qubits rather than qu{\em dits}).  This is largely a matter of convenience, but  
is considerably more common in experimental schemes.   
We will therefore first show how to implement universal computation on an 
encoded qubit that is obtained by discarding the level $|2_L\ra$ in  
Eq. (\ref{eq:qutrit}). Since this encoded qubit is obtained by 
discarding a state from the original qubit, we shall refer to it as a  
truncated qubit. We shall then show how the XY implementation works on 
the full qutrit.  In both cases we are constructing encoded single-qudit 
operations and encoded two-qudit operations that have the capability of 
entangling the encoded logical states, where the qudits are qubits in one 
instance and qutrits in the other.  Thus we are implicitly working within 
the model of encoded universality that provides a map onto the standard model 
of one- and two-qubit operations.

\section{The gate-sequences}  
  
\subsection{Truncated qubit encoding}  
  
We now show how to emulate both single qubit operations and the encoded  
controlled phase-flip. The controlled phase-flip ${ C_z}$, defined by  
$|a\rangle|b\rangle \rightarrow (-1)^{ab} |a\rangle|b\rangle$, is equivalent to  
the controlled-NOT (CNOT), up to local unitary operations on the encoded qubits.  
Note that in the particular encoding we have chosen by discarding $|2_L\ra$  
({\em i.e.}, $|0_L\ra=|100\ra$ and  $|1_L\ra=|010\ra$), the third physical qubit is always  
in the state $|0\ra$. Consequently the third qubit is always $|0\ra$ for any  
state $|\Psi_L\ra=\alpha |0_L\ra + \beta |1_L\ra$ of the encoded qubit.  It is  
therefore redundant for this truncated encoding. However, as we show below, it  
is needed during the application of the gate sequence.  We can say that the {\em  
effective } encoded qubit is placed into a space of only two physical qubits ($|0_L\ra=|10\ra$ and $|1_L\ra=|01\ra$),  
with the third qubit playing the role of an ancilla state (we will need it  
during the application of the gate sequence), which can be reused for several  
computations. We will denote the Pauli-matrices $\sigma_{x,y,z}$ by $X,Y,Z$  
respectively. For notational simplicity, we shall 
subsume the amplitude factors of $J_{ij}$ into the gate phases $\phi$.
Note that the amplitudes $J_{ij}$ are in general allowed to be different for 
different qubit pairs $i, j$. 
  
\paragraph{Single-qubit operations}: We will use the Euler-angle decomposition  
of matrices $U \in SU(2)$ as  
\be  
U=e^{i \phi_1 X} \cdot e^{i \phi_2 Z} \cdot e^{i \phi_3 X}\label{eq:euler}  
\ee  
to show how to implement each of these factors in turn on the encoded qubit.  
This is very easy for the encoded operations generated by $X$:  
\be \label{eq:sigmax}  
(e^{i \phi X})_L=e^{i \phi A_{12}},  
\ee  
where $A_{ij}$ is defined in Eq.~(\ref{eq:H_xy}), {i.e.},  
\be  
A_{ij}= \frac{1}{2} (X^i X^j + Y^i Y^j)  
\label{eq:A}  
\ee  
For the encoded $Z$ we will adapt a sequence from Lidar and Wu \cite{Lidar:01a},  
which acts on three qubits. This sequence, which we will call ${\mathcal P}_3$, will  
be used also to implement the controlled phase-flip $C_z$ between two encoded  
qubits. Its layout is shown in Figure 1:  
\be \label{eq:sigmaz}  
{\mathcal P}_3(\phi)=e^{i \frac{\pi}{4} A_{12}}\cdot e^{i \frac{\pi}{2} A_{23}}\cdot  
e^{i \frac{\phi}{2} A_{13}} \cdot e^{-i \frac{\pi}{2} A_{23}} \cdot e^{-i  
\frac{\pi}{4} A_{12}}.  
\ee  
Figure 2a shows how to use this 5-gate sequence to implement an encoded   
$(e^{i \phi Z})_L$ on our qubit.  
Adding a third ancillary physical qubit in the state $|0\ra$ thus allows us  
to enact ${\mathcal P}_3$ on the truncated qubit $|0_L\ra, |1_L\ra$. In particular,  
${\mathcal P}_3(-\phi)$ transforms $|100\ra \rightarrow e^{i\frac{\phi}{2}}|100\ra$  
and $|010\ra \rightarrow e^{-i \frac{\phi}{2}}|010\ra$.   
This implements the encoded $e^{i \phi Z}$, up to a global phase ($e^{i  
\frac{\phi}{2}}$), and leaves the third qubit unchanged.  
  
Using these two operations, $e^{i\phi A_{12}}$ and ${\mathcal P}_3 $,   
Eq.~(\ref{eq:euler}) shows that we can implement any single qubit gate on the  
encoded  
qubit using at most $7$ gates generated by the XY-interaction.   
  
\paragraph{Two qubit gate}:  
To obtain an encoded ${ C_z}$ between the encoded qubits $|0_L\ra$ and  
$|1_L\ra$ we do not have to work much harder.  
We will demonstrate that the previous $5$-gate sequence ${\mathcal P}_3 $ can be used  
to obtain${ \sqrt{-ZZ}}=\exp(-i \frac{\pi}{4} ZZ)$.  
This gate is equivalent to the controlled phase-flip ${ C_z}$ up to single  
qubit unitaries (${ C_z} = [\exp(i \frac{\pi}{4} Z_1) \otimes \exp(i  
\frac{\pi}{4} Z_2)] \cdot { \sqrt{-ZZ}}$).  
Figure 2b shows two possible layouts of ${\mathcal P}_3 $ that can achieve this, 
{\em i.e.},  
\be  
({ \sqrt{-ZZ}})_L={\mathcal P}_3 (-\frac{\pi}{2})_{123}={\mathcal P}_3 (\frac{\pi}{2})_{124}. \label{eq:sqrtzz}  
\ee  
The subscripts refer to the physical qubits to which ${\mathcal P}_3 $ is applied. This  
means that in order to implement the controlled phase-flip ${ C_z}$ exactly  
(up to single qubit operations), we need only $5$ gates.   
  
\subsection{Universal computation on qutrits} 
Let us now analyze the case of the encoded qu{\em trit}, {\em i.e.}, the three- 
state system of Eq. (\ref{eq:qutrit}). To universally compute on a qutrit, we  
have to show how to implement $SU(3)$ operations on the qutrit and also how to  
make an entangling gate between two encoded qutrits.  As noted above, 
this allows us to use the encoded qutrits to implement quantum circuits 
constructed to implement computation on qutrits. 
  
For the single qutrit operations, we note that it is sufficient to show how to implement $SU(2)$ on each pair of {\em two} of the three states. 
This follows from the enlarging lemma proven in Ref.~\cite{Kempe:00a}.  
Note that both the $X$-gate of Eq. (\ref{eq:sigmax}) and the $Z$-gate of 
Eq. (\ref{eq:sigmaz}) leave the state $|2_L\ra=|001\ra$ unchanged. 
To implement an encoded $e^{i \phi X}$ on the first two states of the qutrit 
we therefore can just apply the $X$-gate of Eq. (\ref{eq:sigmax}) without 
modification. To obtain the $e^{i \phi Z}$ gate we need to be slightly more 
cautious. 
Direct application of the $Z$ gate in Eq. (\ref{eq:sigmaz}) 
transforms $|100\ra \longrightarrow e^{i \phi/2} |100\ra$, 
$|010\ra \longrightarrow e^{-i \phi/2} |010\ra$, but leaves the third state
unchanged, {\em i.e.}, $|001\ra \longrightarrow |001\ra$.  This introduces 
not only the desired relative phase shift between $|0_L\ra$ and $|1_L\ra$,
but also an undesired phase shift between each of these and the third
state $|2_L\ra$.  However, the action of a $Z$ gate on a qutrit should 
introduce only one relative phase, {\em e.g.}, between $|0_L\ra$ and 
$|1_L\ra$, although it can result in an additional global phase. 
To overcome this problem, we can apply the ${\mathcal P}_3 $ gate twice, in an 
arrangement shown in Figure 3a. This now does correctly implement the $Z$ 
gate on the first two states of the qutrit, while maintaining a constant
phase between states $|0_L\ra$ and $|2_L\ra$ and causing an acceptable 
global phase of $e^{i\phi/3}$.  
We can therefore generate $SU(2)$ on $|0_L\ra,|1_L\ra$ without changing 
$|2_L\ra$ with at most $12$ gates ($1$ from each $X$ gate and $10$ from the pair of ${\mathcal P}_3 $ gates). 
By switching qubit $1$ and $3$, these same operations give $SU(2)$ on 
$|1_L\ra,|2_L\ra$. By switching $2$ and $3$, they give $SU(2)$ on $|0_L\ra,|2_L\ra$. 
Hence, by the enlarging lemma \cite{Kempe:00a}, we can use these sequences to
construct any $SU(3)$ operation on the qutrit. 
  
To implement an entangling operation between the two qutrits, we need only to  
implement an entangling operation (e.g., ${ C_z}$) between two of the three  
levels of each qutrit. Using the gates in Eq. (\ref{eq:sqrtzz}), together with  
the fact that ${\mathcal P}_3 $ does not change states of the form $|000\ra$ and  
$|001\ra$, we can implement an encoded ${ \sqrt{-ZZ}}$ between $\{|0_L\ra  
|0_L\ra,|0_L\ra |1_L\ra,|1_L\ra |0_L\ra,|1_L\ra |1_L\ra\}$.   
We achieve this by combining the two options in Eq. (\ref{eq:sqrtzz}) in  
serial, to build a circuit having two ${\mathcal P}_3 $ gates in sequence,  
shown in Figure 3b. We note that the sequence of the two ${\mathcal P}_3 $ gates  
results in a total of only $8$ exchange interactions (rather than $10$), because the last interaction in the first $({\mathcal P}_3) _{123}$ cancels the first one in 
the second $({\mathcal P}_3 )_{124}$. 
We present a second possibility for the same circuit in the lower part of 
Figure 3b. Instead of 
applying the ${\mathcal P}_3 $ gate between qubits $123$ and $124$ in succession, 
we limit the 
application of ${\mathcal P}_3 $ here to just the first three qubits ($123$) in both
instances.  Then we insert an 
XY-gate between qubit $3$ and $4$, which is then undone after the second
${\mathcal P}_3 $ gate. 
This arrangement has two more gates ($10$ instead of $8$), but makes up for 
this by presenting advantages for conception of the layout of the physical 
qubits. The ${\mathcal P}_3 $ gate requires interactions between all three of its 
qubits, which therefore have to be arranged in a geometry which will allow 
for this ({\em e.g.}, triangular). It can therefore be advantageous to limit 
the number of different ${\mathcal P}_3 $ gates.
Yet another alternative to this last circuit is to make the ${\mathcal P}_3 $ gate 
act both times between 
qubits 1,2 and 4, and to insert two XY interactions before and after the first ${\mathcal P}_3 $ gate. 

Whichever of these three entangling ${ \sqrt{-ZZ}}$ circuits is
used, symmetric permutations of the qubits (within one encoded qutrit) will
again allow one to then implement ${C_z}$ between any choice of pairs of two levels of the  encoded qutrits.

\section{State preparation and measurement}

To implement a full-fledged quantum computation with the XY-interaction alone, we have to show how to prepare fresh input states (in the encoded $|0_L\ra$ state) and how to read out the result of the quantum circuit. This task is made easier by the fact that the encoded logical qubit states are all tensor product states of the physical qubits. 

A quantum computation in our scheme would begin by settling all the computational qubits to the $|0_L\ra$ state. In the encoding with three physical qubits this state is of the form $|100\ra$. In the case of a spin-architecture, for instance, single qubits in the $|1\ra$ state can be obtained by placing them in a moderately strong magnetic field pointing in one direction, and the $|0\ra$ spins can be obtained by imposing a magnetic field in the opposite direction. If it is hard to apply this magnetic field locally (such that it affects the first spin only) one may apply the magnetic field to two groups of spatially separated spins to produce separate groups of $|1\ra$ and $|0\ra$ spins. We can then bring these groups of spins close together and use XY-interactions between members of the two groups to ``shift'' the $|1\ra$ spins to the appropriate positions within the other group of $|0\ra$ spins. Note that in order to ``shift'' a $|1\ra$ from, say, position $1$ to position $2$ (up to a global phase) we need only to apply the exchange operator $A_{12}$ for a time $\pi/4$ (see Eq. (\ref{eq:sigmax})). 

To {\em measure} the outcome of a computation we only need to distinguish $|0_L\ra$ from $|1_L\ra$ (or from $|2_L\ra$ if we use all three qutrit states for encoding). This means a measurement has to determine whether the first physical qubit is in the state $|1\ra$ or $|0\ra$. This can again be done by preparing a (spatially separated) group of spins in the $|0\ra$-state and using the XY-interaction to ``shift'' the qubit state in question into this group. A (destructive) measurement of the total magnetic moment can then determine whether the spin in question was in the state $|0\ra$ or $|1\ra$.

These elements of state preparation and measurement complete the set of primitives needed for fault-tolerant quantum computation.

\section{Layout}  
  
The gate operations needed for univeral computation with the XY-interaction 
only will not necessarily involve just nearest neighbor interactions. 
In fact,  
it can be shown that in a linear array, nearest neighbor XY interactions 
alone are not universal  
with any encoding (see \cite{Kempe:01a} and references therein). A layout of 
the physical qubits has to respect this fact, introducing issues of 
architecture into consideration. In Figure 4 we present two possible  
two-dimensional layouts, one for encoded truncated qubits (a) and the other 
for the full qutrits (b). The triangular arrangement allows for all possible  
interactions between the three qubits that make up an encoded truncated qubit 
or an encoded qutrit, using only nearest neighbor interactions. 
This allows 
application of the ${\mathcal P}_3 $ gate on the three qubits of the triangle using
only nearest neighbor interactions. The 
arrangement of the triangles is such, that there is always a triangular shape 
between two qubits of one triangle and one qubit of the next. 
This way we can always apply a ${\mathcal P}_3 $ gate between two qubits of one encoded qutrit and one qubit of the other, 
which is sufficient to implement the entangling gates also using only
nearest neighbor interactions. 
Note that in the case of Figure 3a) we might need to apply XY-interactions 
between the two physical qubits that encode a truncated qubit and the third (ancillary) qubit, set to $|0\ra$, before applying the ${\mathcal P}_3 $ gates for the two-qubit operations.

Within this architecture, the fundamental interaction 
is indeed simply a nearest neighbor interaction.  In fact, it is not hard to  
see  
that in a linear array, nearest and next-nearest neighbor XY-interactions 
will suffice for universal computation. 
Any  
layout that allows for these two types of interactions will therefore
allow for universal  computation. The two-dimensional triangular layout proposed here has the 
advantage of requiring just nearest-neighbor interactions.  This can be
either symmetric, ({\em i.e.}, $J_{ij} \equiv J$ for all $i,j$), or
inequivalent, without modification of the present gate sequences.
Note that in the case of the truncated qubit we only need a third ancillary 
qubit (set to $|0\ra$) whenever a $Z$-gate 
(Eq. (\ref{eq:sigmaz})) is to be performed. The effective encoding is thus
$|0_L\ra=|10\ra$ and $|1_L\ra=|01\ra$. 
One can imagine alternative arrangements that supply the
ancillary qubit in the 
state $|0\ra$, either permanently or in a temporary fashion whenever a $Z$ gate needs to be performed. A three-dimensional 
architecture, such as that presented in Figure 5, can achieve this with
an economy of qubits resulting from a mobile ancilla scheme. 
Here each truncated qubit is encoded by two physical 
qubits (all in one plane) and the single ancillary qubit is moved around in
a second plane, to be used whenever needed for implementation of a $Z$ gate.
If such ancilla qubit mobility is not practical, one simply uses a static row 
of ancillary qubits in the second plane.

The XY exchange interaction underlies several experimental 
proposals for solid-state qubits, including quantum dot spins coupled
by cavity QED \cite{Imamoglu:99} and nuclear spins coupled by a 
two-dimensional electron gas \cite{Mozyrsky:01}. These different
proposals involve diverse requirements on realizing physical coupling of
the qubits.  For the quantum dot/cavity QED proposal, a two-dimensional
triangular layout such as those in Figure 4 is attractive, while the
three-dimensional layout of Figure 5 (with a static array of ancilla qubits
in the second plane) appears well suited to the long-range
coupling of nuclear spins proposed in Ref. \cite{Mozyrsky:01}. Clearly, the  
ultimate layout in physical implementations will be subject also
to experimental restrictions on the particular physical 
system at hand.  
  
\section{Conclusion}

We have constructed efficient gate sequences to implement universal
quantum computation using the anisotropic exchange (XY) interaction
alone.  The most compact solution allows definition of qubits as truncated
qutrits, and results in quantum logic elements containing at most $7$ XY
exchange
operations for encoded single qubit operations, and a maximum of $5$ XY
operations for encoded two-qubit gates.  These explicit gate sequences
offer an attractive route to implementation of quantum information
processing using transverse spin-spin interactions. A prototype two-dimensional layout was
suggested here that lends itself well to the architecture of quantum dots
coupled by cavity QED proposed in Ref.~\cite{Imamoglu:99}.

\vskip1cm 
{\bf Acknowledgments}: JK and KBW's effort is sponsored by the Defense Advanced  
Research Projects Agency (DARPA) and Air Force Laboratory, Air Force Material  
Command, USAF, under agreement number F30602-01-2-0524 and FDN00014-01-1-0826. 

\bibliographystyle{prsty}  
  

\pagebreak  
  
\vbox{  
\begin{figure}\label{fig1}  
\epsfxsize=14cm  
\epsfbox{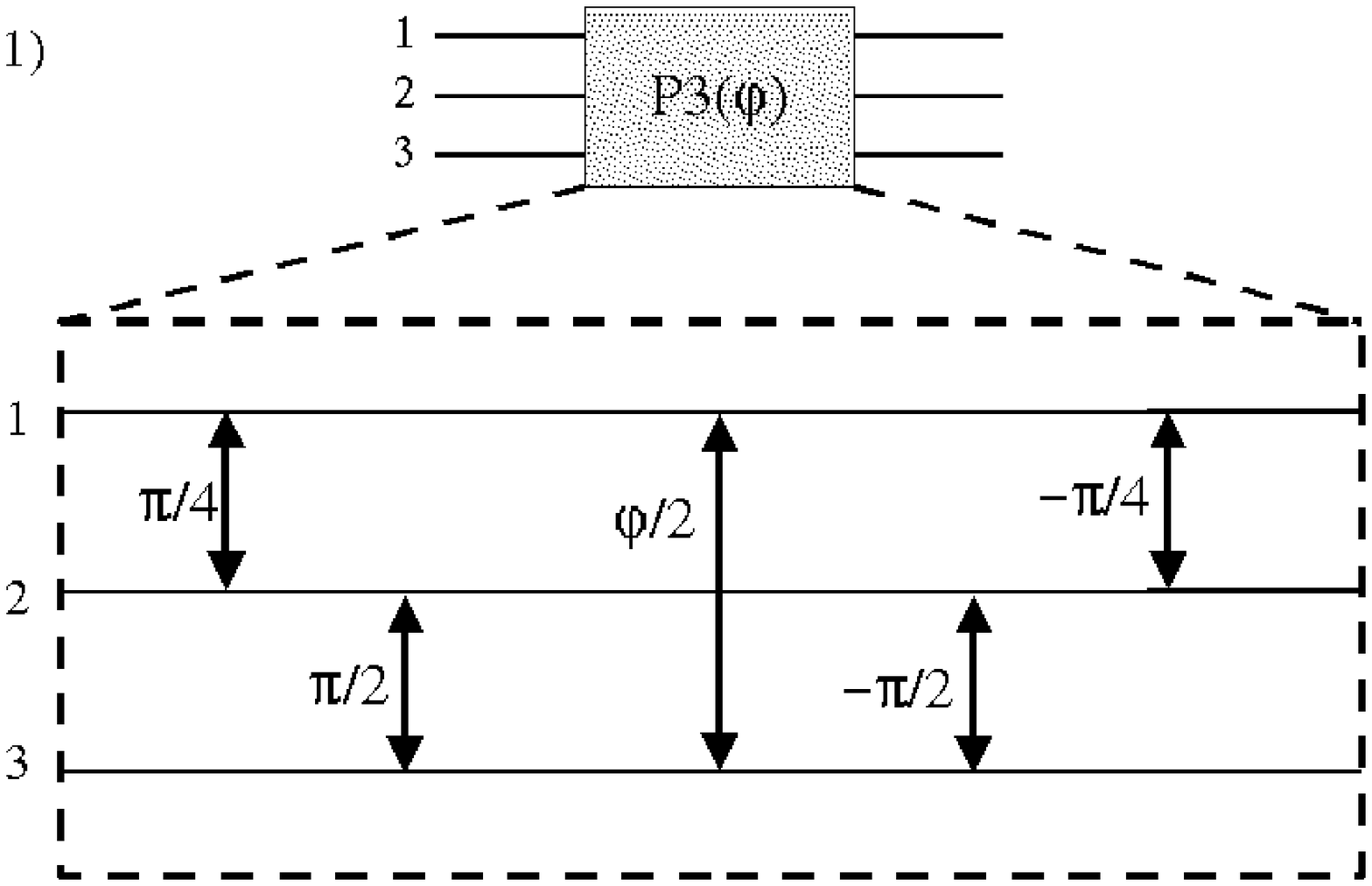}  
\smallskip  
\caption{Three-qubit ${\mathcal P}_3 $ gate. The diagram depicts the layout of the five  
XY-interactions to be applied to three physical qubits (denoted 1, 2, and 3).  
The time lines of the qubits run from left to right. The arrows represent the  
XY-interactions with labels showing the corresponding interaction times. The  
time of the central XY-interaction, $\phi/2$, will be set according to the 
final use of ${\mathcal P}_3 $, so we refer to the gate as ${\mathcal P}_3 (\phi)$.  This 
network transforms $4$ of the $8$ total states of the three qubits, according 
to    
$|0,1,0\ra \rightarrow  e^{i \frac{\phi}{2}} |0,1,0\ra$ ,   
$|1,0,0\ra \rightarrow  e^{-i \frac{\phi}{2}} |1,0,0\ra$,   
$|0,1,1\ra \rightarrow  e^{-i \frac{\phi}{2}} |0,1,1\ra$,   
$|1,0,1\ra \rightarrow  e^{i \frac{\phi}{2}} |1,0,1\ra$.  It leaves the  
remaining $4$ states unchanged.  
}  
\end{figure}  
}  
  
\vbox{  
\begin{figure}\label{fig2}  
\epsfxsize=14cm  
\epsfbox{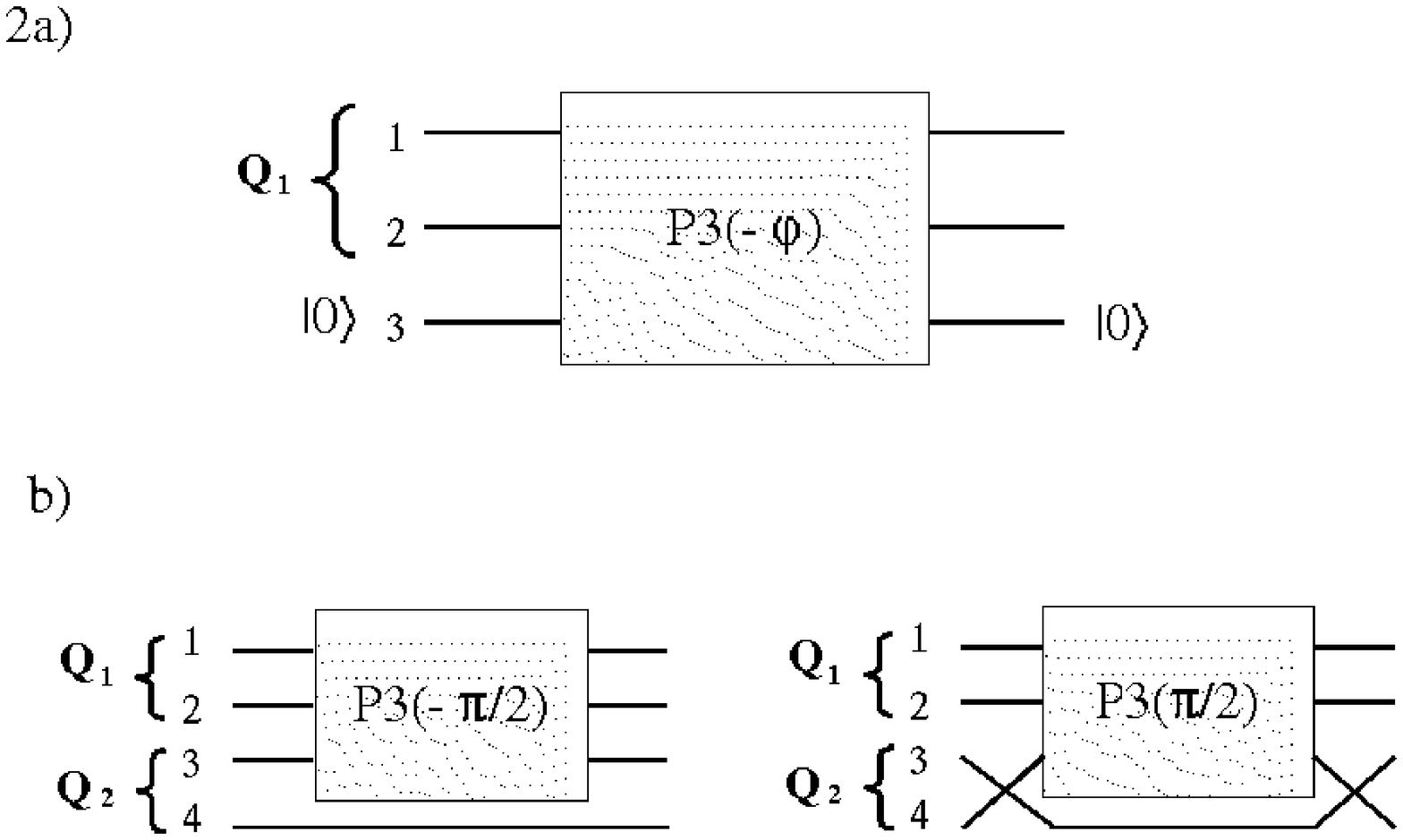}  
\smallskip  
\caption{Quantum logic with the ${\mathcal P}_3 $ gate on encoded truncated qubits.  
{\bf 2.a}  gives an implementation of the single qubit $e^{i \phi Z}$ gate. The  
third (ancillary) qubit is set to $|0\ra$ initially. ${\mathcal P}_3 (-\phi)$  
implements the transformation   
$|1,0,0\ra \rightarrow e^{i \frac{\phi}{2}} |1,0,0\ra$ and $|0,1,0\ra  
\rightarrow  e^{-i \frac{\phi}{2}} |0,1,0\ra$, which corresponds to $e^{i \phi  
Z}$ (up to a global phase). After the gate operation the third qubit is once  
again in the $|0\ra$ state and can be reused.   
{\bf 2.b} shows two possible options of how to implement the ${ \sqrt{-ZZ}}$  
gate on two encoded truncated qubits. Here ${ Q_1} \equiv 1, 2$ and ${  
Q_2} \equiv 3, 4$ represent the two first physical qubits of the three qubit  
encoding.  The third, ancillary, qubits of each truncated qubit are not involved  
in these gate sequences and so are not shown here.     
Both circuits transform   
$|1,0,1,0\ra \rightarrow  e^{-i \frac{\pi}{4}} |1,0,1,0\ra$,  
$|1,0,0,1\ra \rightarrow  e^{i \frac{\pi}{4}} |1,0,0,1\ra$,   
$|0,1,1,0\ra \rightarrow  e^{i \frac{\pi}{4}} |0,1,1,0\ra$,  
$|0,1,0,1\ra \rightarrow  e^{-i \frac{\pi}{4}} |0,1,0,1\ra$, which is equivalent  
to the logical ${ \sqrt{-ZZ}}$ on the logical states $\{|0_L\ra  
|0_L\ra,|0_L\ra |1_L\ra,|1_L\ra |0_L\ra,|1_L\ra |1_L\ra\}$ of the encoded  
truncated qubits. }
\end{figure}  
}  
\vbox{  
\begin{figure}\label{fig2c}  
\epsfxsize=14cm  
\epsfbox{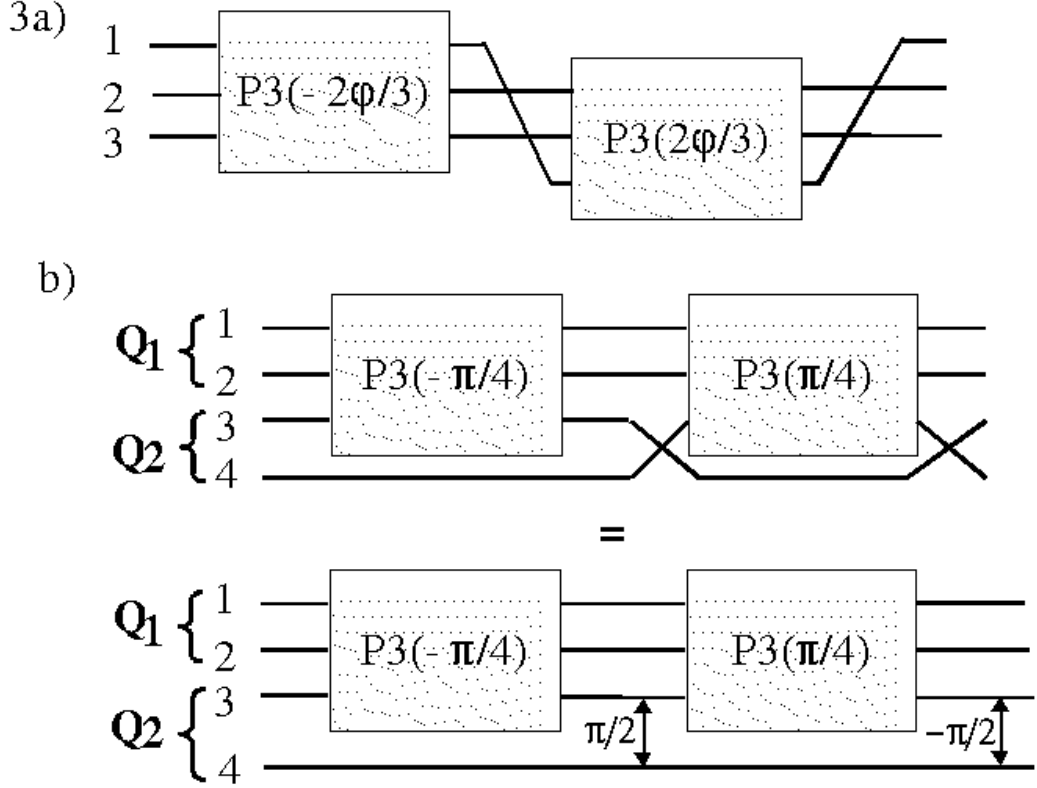}  
\smallskip  
\caption{Quantum logic with the ${\mathcal P}_3 $ gate on encoded qutrits.
{\bf 3.a} shows how to use two ${\mathcal P}_3 $ gates to implement $e^{i \phi Z}$ on the first two levels of an encoded qutrit. The net transformation is $|1,0,0\ra \rightarrow e^{i \frac{\phi}{3}} |1,0,0\ra$, $|0,1,0\ra \rightarrow  e^{-i \frac{2 \phi}{3}} |0,1,0\ra$ and $|0,0,1\ra \rightarrow  e^{i \frac{\phi}{3}} |0,0,1\ra$, which corresponds to the $Z$ gate on the first two states up to a global phase of $e^{i \frac{\phi}{3}}$. {\bf 3.b} shows two equivalent ways of how to use two ${\mathcal P}_3 $ gates to implement ${ \sqrt{-ZZ}}$  
on the states $\{|0_L\ra |0_L\ra,|0_L\ra |1_L\ra,|1_L\ra |0_L\ra,|1_L\ra  
|1_L\ra\}$ of an encoded qutrit, while leaving all other states unchanged. 
Here  ${ Q_1} \equiv 1, 2$ and ${ Q_2} \equiv 3, 4$ represent the two 
first physical qubits of the three qubit encoding.  The third qubit of  
each qutrit is not involved in these gate sequences and so is not shown here. The lower gate sequence has two more gates, but has the advantage that it can be implemented in the triangular arrangement of Figure 4b.} 
\end{figure}  
}  
\newpage
\vbox{  
\begin{figure}\label{fig3}  
\epsfxsize=14cm  
\epsfbox{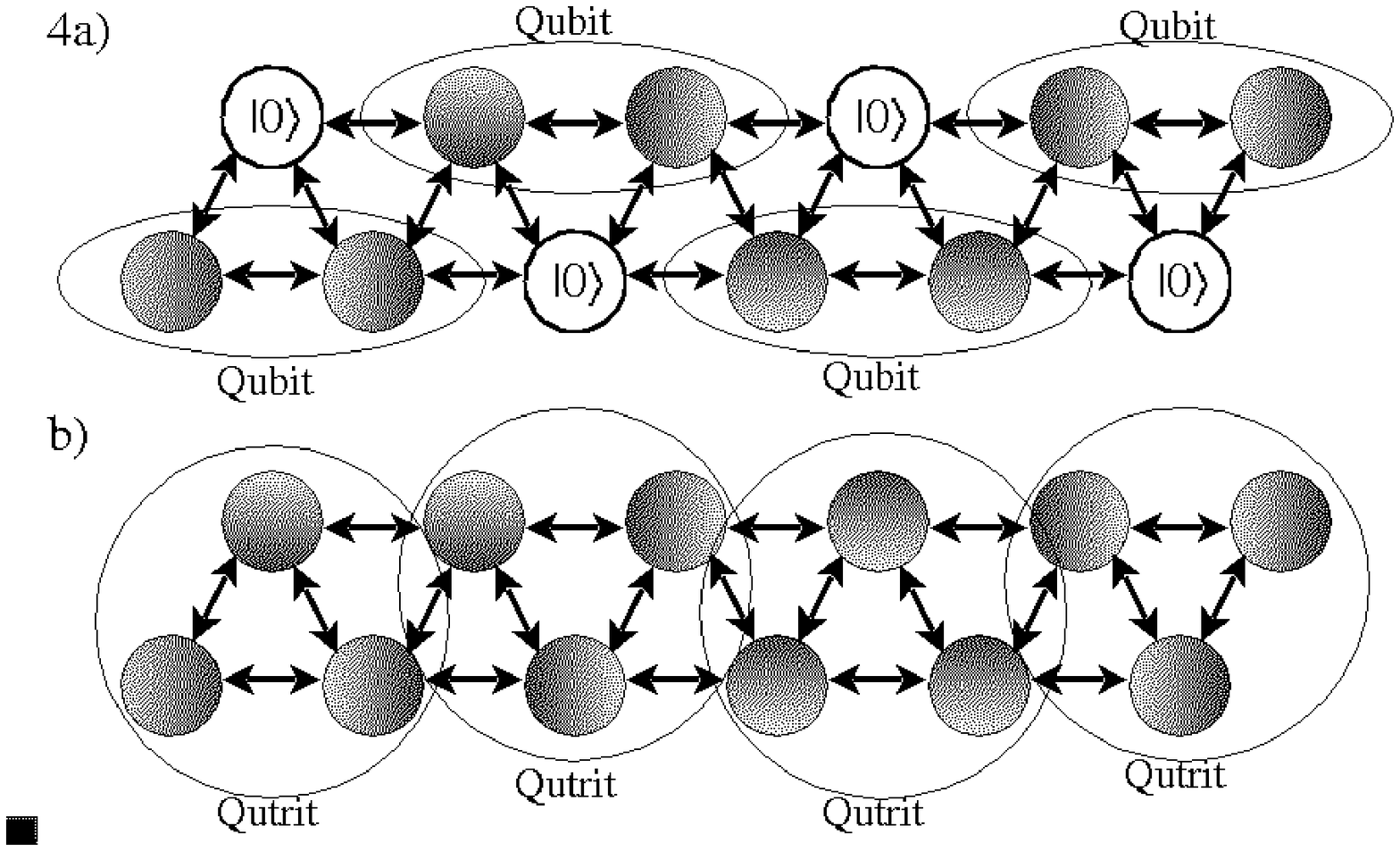}  
\smallskip  
\caption{Possible layouts for encoded qubits or qutrits. In {\bf 4.a} we give a  
possibility to arrange the physical qubits in two lines. The arrows represent  
controllable $XY$-interactions. For each truncated qubit we have an additional  
ancillary qubit set to $|0\ra$ that is needed to perform the encoded $e^{i \phi  
Z}$. In {\bf 4.b} we show how this same array can be used for encoding  
three qubits into full qutrits in triangular arrangement. }  
\end{figure}  
}  
\newpage
\vbox{  
\begin{figure}\label{fig5}  
\epsfxsize=14cm  
\epsfbox{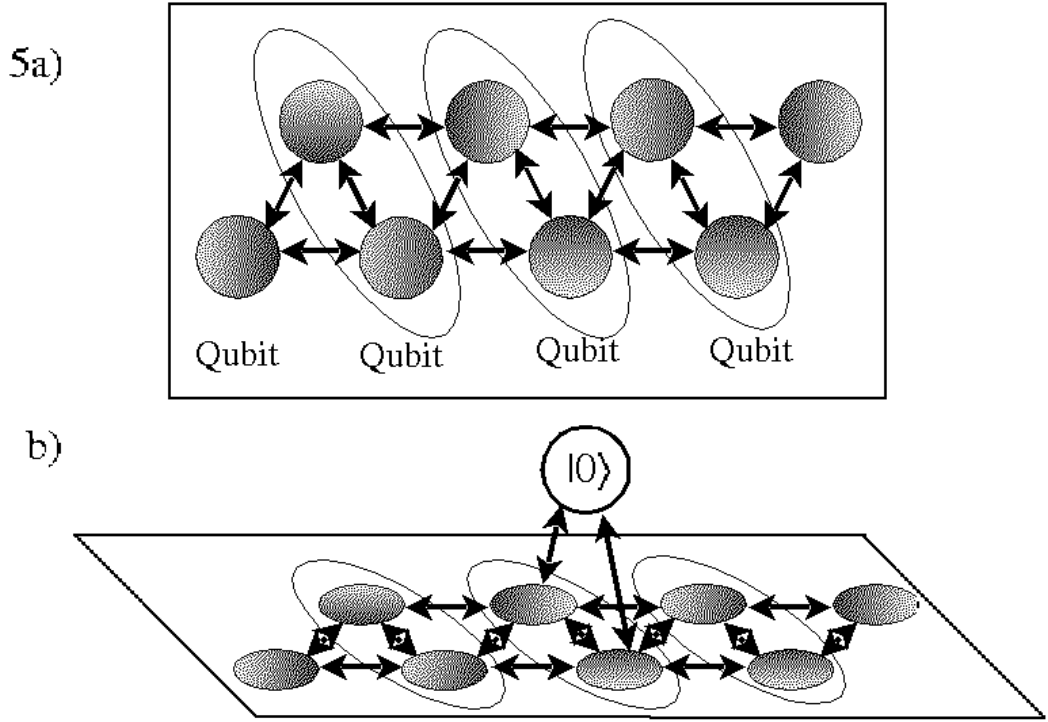}  
\smallskip  
\caption{Three-dimensional layout for truncated qubits. {\bf 5a)} The encoded qubits $|0_L\ra=|10\ra$ and $|1_L\ra=|01\ra$ are arranged in pairs in one
plane.  {\bf 5b)} A third, ancillary, 
qubit (needed for the $Z$ gate) is located in the plane above the qubits. 
The third qubit may be a member of a stationary array of ancilla qubits, configured such 
that each pair of qubits has access to one ancilla.
Mobile ancilla qubits offer an economy arrangement, in which a single ancilla 
is transported to the location where a $Z$ gate is to be performed at each 
time step.
 }  
\end{figure}  
}  
  
\end{document}